\documentclass[twocolumn,showpacs,preprintnumbers,amsmath,amssymb]{revtex4}

\usepackage{graphicx}

\begin{document}


\title{Explaining the pearl necklace of SNR 1987A by coherent optics.}

\author{Jacques Moret-Bailly}
\affiliation{Physique, Universite de Bourgogne Dijon France}
\email{jacques.moret-bailly@u-bourgogne.fr}
\date{\today}

\begin{abstract}
A lot of beautiful observations of Supernova remnant 1987A give a precise idea of its structure and its evolution. The regular interpretations of the observations set that the large energy needed to explain the brightness of the “pearl necklaces” is provided by shock waves involving remnants of a first explosion and a wave produced by the observed explosion although the existence of this wave is discussed. We develop the alternative explanation of the necklaces by photoionization. Our main hypothesis is that the explosion of the blue supergiant progenitor produces two neutron stars and a central brilliant object, a linear system similar to those which were observed by Halton Arp. We suppose that these stars remains bright in extreme UV, to maintain the strong ionization of a bubble of hot hydrogen nearly transparent in far UV ( defined as the range of Lyman frequencies of atomic hydrogen). Outside the bubbles, three shells containing atomic hydrogen generate resonant, superradiant scatterings at Lyman frequencies, in tangential competing modes. The superradiance cools the gas and absorbs strongly the radial far UV light, hiding the stars. The shells may be identified with the inner active shells found from light echoes.
\end{abstract}

\maketitle

\section{Introduction}
Light and neutrinos from Supernova 1987A arrived at Earth on February 23, 1987 after a 166 000 year trip. Its brightening was very rapid: by a factor 100 in 3 hours, reaching magnitude 2,9 in 80 days. This slow rise, and the slow speed of hydrogen observed in the infrared show that the mass of hydrogen envelope of the star was of the order of 10 solar masses. Then the magnitude decreased almost linearly to 16 in 1000 days \cite{Arnett}. Its initial spectrum was very rich in UV and showed broad hydrogen and helium emission lines; then the UV decreased and it remained only emission lines of low ionization elements \cite{Woosley}.
Ground based images of SN1987A showed a weak blob of gas which was resolved as a circumstellar  “equatorial ring” (ER) by the ESA Faint Object Camera on Hubble on August 1990. Comparing the time delay between the maximal emissions of the supernova and the ring gave a diameter of the ring: 1.2 light-year (ly); knowing its angular size, the distance of the supernova was set \cite{Panagia}.
In 1994, two larger “outer rings” (OR) were detected; in 1997, bright spots (pearls) appeared on the ER.
The evolution of the rings (necklaces), the existence of the pearls is now explained using shock waves between matter ejected by the supernova and matter ejected 20 000 years before in a first explosion of a red supergiant star; however, arguments developed for instance by LLoyd et al. \cite{Lloyd}, such as the complexity of the explanation of the existence of the pearls, of the thinness of the rings, their inner brightness and of the disappearance of the star let us suggest the use of an alternative explanation closer to the interpretation of the spectra of planetary nebulae \cite{Plait}.

Our fundamental hypothesis, developed in section \ref{SGS} is that the result of the explosion of the red supergiant progenitor star is not only the blue supergiant which was observed, but, also, two neutron stars aligned  with it on a z' axis. Inside the rings, we follow the photoionization model developed by Chevalier et al. \cite{Chevalier} and Lundqvist et al. \cite{Fransson,Lundqvist}.

Section \ref{in} explains the emission and diffusion of light inside the ER and the observation of anomalous frequency shifts.

Section \ref{ER} describes the generation of the equatorial ring.

Section \ref{Geo} studies the geometry of the rings.

{\it All ellipsoids used in this paper are invariant by rotations around axis z'.}

\section{The supergiant progenitors.}\label{SGS}
The explosion of the red supergiant progenitor 20 000 years before the observed blue supernova explosion \cite{Panagia}, results from a collapse of its core by a fusion of the protons of iron nuclei with electrons resulting from a cooling of this core. It is usually admitted that this collapse produces a prompt shock, and that the shock wave may propagate outside the star. However this process works only with a small core \cite{Myra} while the mass of the star is large, at least 20 times the mass of the Sun.

 Halton Arp observed a number of alignments of quasars with a larger object often similar to a BL Lac, too large to be accidental \cite{Arp}. These alignments seem result from the explosion of a progenitor of the larger object. The good alignment of three objects seems indicate the following scenario: the conservation of the angular momentum during the collapse of the core of the progenitor increases the angular speed of the core so much that, like a spinning drop of liquid, it increases its momentum of inertia, having successively the shapes of an oblate ellipsoid, a prolate ellipsoid, a peanut, finally exploding. Whichever the exact process, we suppose that the explosion of the red progenitor of SN 1987A produced, along a z' axis, the blue observed supergiant and two neutron stars; this system is surrounded by ejected external layers of the red supergiant, having the third of its mass, and a density decreasing with the distances to the supergiant star and to z' axis. This cloud of gas is mainly made of hydrogen and helium. The bipolar symmetry of the cloud was observed by Wang et al. \cite{Wang} and deduced by Sugerman et al. from the observation of light echoes \cite{Sugerman}.

Accreting the cloud, the neutron stars play the role of the anticathode of an X-rays tube (except for the origin of the acceleration of the particles). Thus, the hot spots emit X rays and extreme UV, but as they are very small, in despite of a temperature of several millions of kelvins, their total radiation at lower frequencies is too weak to be observable in optical regions. The blue supergiant is heated both by the accretion and its internal evolution; its emission in extreme UV is not much larger than the emission of the neutron stars, but it is much more powerful in far UV region, around the Lyman lines of hydrogen.

\section{The plasma inside the equatorial ring (ER).}\label{in}
After the blue supernova explosion, the remaining star had the spectrum of a blue star, emitting very broad Lyman lines of H$_{I}$ and He, and a strong continuous spectrum of shorter fréquencies (extreme UV). We suppose that, after the main supernova glow, the surface temperature of the star does not decrease very strongly, so that its spectrum remains mainly made of less wide Lyman lines of H$_{I}$ and extreme UV. In a first approximation, suppose that the whole system has a spherical symmetry. The cooling of the plasma is limited by the absorption of the extreme UV which ionizes generated H$_{I}$ into electrons and protons (H$_{II}$), building a spherical bubble of H$_{II}$ inside a cloud of H$_{I}$ \cite{Chevalier}. In the bubble, various ionized atoms (He, C, N, O, ...) absorb some energy, radiate lines and produce a Rayleigh scattering of the Lyman emission lines. The largest fraction of this scattering is coherent, producing only the refraction; the weaker incoherent scattering produces an opalescence around the star, similar to the blue of our sky. As protons and electrons are nearly free, they do not play a notable optical role. As in the atmosphere of the Earth, the Raman incoherent scattering is weaker than the Rayleigh incoherent scattering.

\begin{figure}
\includegraphics[height=6 cm]{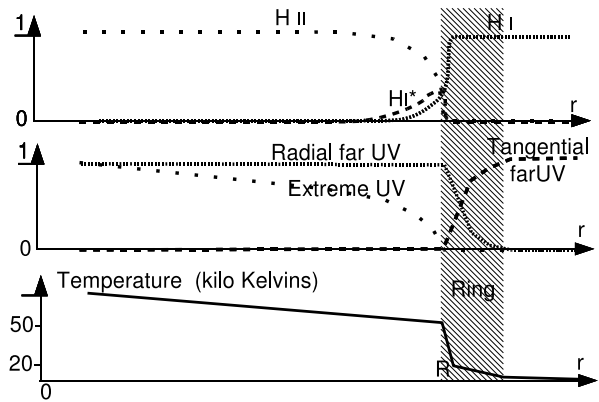}
\caption{\label{rad} Variation of the relative densities of H$_I$, H$_{II}$ and excited atomic hydrogen H$_I$*, relative intensities of light, and temperature along a radius of the equatorial ring.}
\end{figure}

\begin{figure}
\includegraphics[height=6 cm]{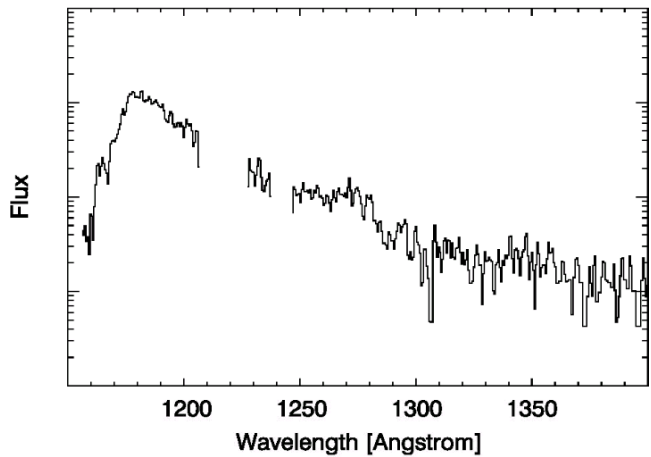}
\caption{\label{foret} Observed profile from one spatial region. Blank regions are due to contaminants. From Michael et al. \cite{Michael}}
\end{figure}

For a chosen orientation, locate a point by its distance r to the centre of the star, and set r = R at the inner point of the necklace (figure \ref{rad}). For r larger than about 3R/4, the remaining intensity in extreme UV has decreased enough, and the ionization by several steps is improbable enough, to let appear some H$_{I}$, with a density increasing with r, so that the broad Lyman lines are scattered more strongly \cite{Sonneborn}. Very close to the ring, an anomalous redshift of the Lyman alpha line appears, see, for instance Lawrence \& Crotts \cite{Lawrence}; a wide absorption band, probably resolvable into lines was observed by Michael et al. \cite{Michael}, similar to the \textquotedblleft Lyman forest'' observed in the spectra of the quasars (fig. \ref{foret}). An explanation, weakly supported by its authors, is a Wolf process, multiple incoherent scatterings combined with an asymmetrical weakening of the aisles of the lines \cite{Wolf}.

 This incoherent process is too weak to allow the large redshifts which are observed, it blurs the images and the spectra too much. An \textquotedblleft Impulsive Stimulated Raman Scattering'' (ISRS) \cite{{Yan,Weiner,Dougherty,Dhar}} may occur because, in a low pressure gas containing H$_{I}$ in 2s or 2p states, the conditions set by G. L. Lamb \cite{Lamb} are fulfilled: the nanosecond pulses which make the ordinary incoherent light are \textquotedblleft shorter than all relevant time constants'', the collisional time, and Raman resonance periods corresponding to 178 MHz in the 2s$_{1/2}$ state, 59 MHz in 2p$_{1/2}$ state, and 24 MHz in 2p$_{3/2}$. 

The ISRS renamed \textquotedblleft Coherent Raman Effect on Incoherent Light'' (CREIL) when it uses ordinary incoherent light rather than ultrashort laser pulses, increases the entropy of a set of electromagnetic beams propagating in a medium verifying Lamb's conditions, by frequency shifts; it does not blur the images because it is coherent, or the spectra because the relative frequency shifts are nearly constants. Usually, light is redshifted, radiofrequencies are blueshifted (in particular, the thermal background is heated).

The large number of lines which appear in a Lyman forest results from the following process in unexcited H$_{I}$\cite{Mor98b,Mor01}: 

The absorption of the Ly$\alpha$ line produces 2p atoms, therefore a CREIL frequency shift; suppose that a partial absorption shifts the absorbed frequency out of the bandwidth of the Ly$\alpha$ line; the pumping and the shift are permanent until a previously absorbed line reaches the Ly$\alpha$ frequency, so that the frequency shift stops, absorptions and emissions become regular, increasing the number of lines in the spectrum. The shift restarts not easily, mainly by a population of the metastable 2s state resulting from a de-excitation of higher states pumped by other Lyman absorptions. In particular, each shift to the alpha frequency, of Lyman beta and gamma lines written in the spectrum, writes new beta and gamma lines, generating the forest with the characteristic fundamental relative frequency shift 0.062 which is observed in the spectra of the quasars \cite{Karlsson,Mor06}.

\section{Generation of the equatorial ring (ER).}\label{ER}

In despite of the previously described scatterings, far UV light propagates in ionized hydrogen without a large absorption until it reaches a region cold enough to decrease strongly the ionization of hydrogen: under 50 000 K atomic hydrogen is stable, mainly in excited states; this temperature was observed by Arnett et al. close to the necklace \cite{Arnett}.

Containing much hydrogen atoms in the excited states, the medium is able to amplify the Lyman frequencies much; in their ground state, the hydrogen atoms scatter the Lyman lines too; therefore, in directions where the excited medium is thick, that is tangentially to the bubble, a strong superradiance appears, which starts a catastrophic evolution of the gas:

The superradiance depopulates the excited states in a spherical shell, cooling hydrogen strongly; ionization disappears, a large amount of atoms becomes available. A coherent nearly resonant scattering transfers directly energy from the radial modes to the tangential modes, exactly as in a laser medium pumped transversally. If the thermodynamic equilibrium were reached, the luminance of the tangential modes would reach the luminance of the radial modes; but this is not necessary to hide the star because the ratio of the solid angles through which are seen, from the Earth, respectively the star and a pearl of the necklace, is very small. The opalescence is not weakened because its luminance, therefore its temperature computed from Planck's law, is low.

As energy is strongly transferred from the radial rays, the tangential luminance of the shell decreases for r increasing \cite{Plait,Panagia}.

\medskip
\begin{figure}
\includegraphics[height=5 cm]{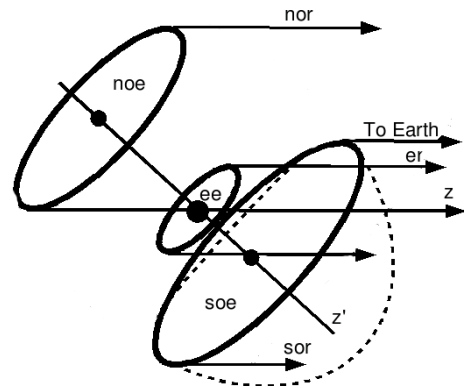}
\caption{\label{fig_1} Section, by the plane containing the axis of symmetry z' and the direction z to Earth, of the ellipsoidal shells of atomic hydrogen which generate the necklaces; noe, ee and soe: north outer, equatorial and south outer ellipsoids; nor, er, sor: traces of the hollow, tangent cylinders of light oriented to Earth; the dotted lines suggest how, taking into account the variations of density of the gas, the outer shells can get a more realistic shape of pear.}
\end{figure}

The tangential rays directed to the Earth draw a circular \textquotedblleft necklace''; but the competition of the modes selects some of them, stabilized by permanent perturbations of the system, for instance by dust: the selected modes make the \textquotedblleft pearls''. Remark that the mathematical definition of the modes does not imply that they are monochromatic: a mode is a ray in the space of the solutions of a linear set of equations (here Maxwell's equations). This is happy because a perfectly monochromatic mode never starts and stops, it is not physics! Here, the nonlinearity of the superradiant process provides an interaction between modes which could be defined for different frequency bands.

The columns of light generated by the superradiant modes excite various atoms, generating columns of excited atoms. These cold atoms radiate nearly monochromatic modes collinear with the initial UV modes, increasing the spectral complexity of the pearls.

\section{Geometry of the rings.}\label{Geo}
We have assumed that the gas is homogeneous to obtain spherical shells; assume now that the density is a decreasing function of r, the surfaces of equal density being prolate ellipsoids whose axis is the z' axis of the stars. The far UV ionizing light is absorbed by a certain column density of H$_{I}$, therefore, for instance, more strongly on z' axis than in other directions; thus, the radius of the shell is lower in this direction than in the other; the shell gets a shape close to an oblate ellipsoid (fig \ref{fig_1}). The cylinder tangent to the ellipsoid, directed to the Earth is elliptical.

\medskip

The accretion of the surrounding cloud of hydrogen by the neutron stars heats very small \textquotedblleft hot spots'' to a temperature larger than 10$^6$ K, so that these stars radiate light mainly in extreme UV. 
This UV generates shells of gas scattering superradiant light beams; these shells are more poorly illuminated than the equatorial shell, by their quasars in far UV and by the columns of light emitted by the ER. The variation of the density of the gas in the z' direction is asymmetric, so that the shell takes the shape of a pear, the external necklaces loosing their centre of symmetry.

\begin{figure}
\includegraphics[height=6 cm]{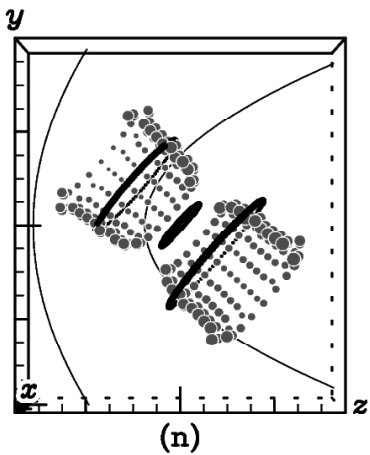}
\caption{\label{fig_2}Section of the \textquotedblleft circumstellar hourglass'' scattering shells found from echoes by Sugerman et al. and the rings (fig. 43n of \cite{Sugerman}). Compare with fig. \ref{fig_1} obtained roughly from our hypothesis.}
\end{figure}

Very similar geometries of light reflecting shells were obtained experimentally by Sugerman et al. \cite{Sugerman} whose figure 43n, reproduced as figure \ref{fig_2} shows a metal cup shape close to our pear shape.

\section{Conclusion}
Several authors tried to explain the generation of the pearl necklaces of supernova remnant 1987A using a photo excitation of surrounding hydrogen: extreme UV maintains the ionization of a bubble an hydrogen cloud. outside of the bubble, a shell of atomic hydrogen scatters far UV light. This explanation fails because it does not explain the fast decrease of the luminance of the ring for an increase of the radius; this failure is removed taking superradiance into account. The existence of the pearls results from a competition of modes as in a laser. The anomalous frequency shifts result from an other coherent effect usually observed using lasers. 

The presence of three rings is explained by a nearly symmetrical ejection of two neutron stars by the red progenitor of the supernova. The accretion of matter by these neutron stars provides the extreme UV needed to generate the scattering shell, while the scattered light is amplified too by light coming from the necklace of the main remnant star.

\end{document}